\documentclass[
 reprint,
 superscriptaddress,
 amsmath,
 amssymb,
 nobibnotes,
 aps,
 prx
]{revtex4-2}

\usepackage{graphicx}
\usepackage{dcolumn}
\usepackage{bm}

\usepackage[english]{babel}
\usepackage{amsmath,graphicx,textcomp}
\usepackage{xcolor}
\usepackage{mathtools}
\usepackage{braket}
\usepackage[normalem]{ulem}

\usepackage{multibib}
\newcites{supp}{Additional References}

\usepackage[colorlinks=true, hyperindex, breaklinks, urlcolor=blue, linkcolor=blue, citecolor=blue]{hyperref}

\definecolor{db}{RGB}{1, 33, 105}
\definecolor{po}{rgb}{1,0.56,0}
\definecolor{yaleblue}{rgb}{0.06, 0.3, 0.57}
\hypersetup{citecolor=db,urlcolor=db}

\newcommand{\ece}{Department of Electrical and Computer Engineering, Princeton University, Princeton, NJ 08544, USA}

\newcommand{\phy}{Department of Physics, Princeton University, Princeton, NJ 08544, USA}

\newcommand{\yale}{Department of Applied Physics, Yale University, New Haven, CT 06520, USA}

\newcommand{\lt}{\tau}
\newcommand{\tsq}{t_{\rm{1Q}}}

\newcommand{\tprep}{t_{\rm{prep}}}
\newcommand{\twait}{t_{\rm{wait}}}
\newcommand{\esq}{\epsilon_{\rm{1Q}}}
\newcommand{\ecz}{\epsilon_{\rm{CZ}}}
\newcommand{\ssz}{$^1\rm{S}_0$~}
\newcommand{\tpz}{$^3\rm{P}_0$~}
\newcommand{\yybb}{$^{171}$Yb~}
\newcommand{\yybbns}{$^{171}$Yb}

\begin{document}

\title{Logical qubits with erasure conversion using metastable neutral atoms}

\author{Bichen Zhang}
\thanks{These authors contributed equally to this work.}
\affiliation{\ece}

\author{Genyue Liu}
\thanks{These authors contributed equally to this work.}
\affiliation{\ece}

\author{Guillaume Bornet}
\affiliation{\ece}

\author{Sebastian P. Horvath}
\affiliation{\ece}

\author{Pai~Peng}
\thanks{Present address: School of Physics, Peking University, Beijing 100871, China.}
\affiliation{\ece}

\author{Shuo Ma}
\thanks{Present address: Department of Physics, University of California, Berkeley, CA 94720, USA.}
\affiliation{\phy}

\author{Shilin Huang}
\thanks{Present address: Department of Physics, The Hong Kong University of Science and Technology, Clear Water Bay, Kowloon, Hong Kong, China.}
\affiliation{\yale}

\author{Shruti Puri}
\affiliation{\yale}

\author{Jeff D. Thompson}
\email{jdthompson@princeton.edu}
\affiliation{\ece}

\begin{abstract}
Implementing large-scale quantum algorithms with practical advantage will require fault-tolerance achieved through quantum error correction, but the associated overhead is prohibitive. This overhead can be reduced by engineering physical qubits with fewer errors, and by shaping the residual errors to be more easily correctable. In this work, we demonstrate quantum error correcting codes and logical qubit circuits in a metastable ytterbium-171 nuclear spin qubit with a noise bias towards erasure errors. These errors can be located separately from any syndrome information diagnosing the error, and we demonstrate adaptive circuit execution based on erasure information. We show that dephasing errors on the qubit during coherent transport can be strongly suppressed, and implement entangling gates that maintain a high fidelity in the presence of gate beam inhomogeneity or pointing errors. Furthermore, we demonstrate logical qubit encoding in the $[[4, 2, 2]]$ code, with error correction during decoding based on mid-circuit erasure measurements despite the fact that the code is too small to correct any Pauli errors. Finally, we demonstrate logical qubit teleportation between multiple code blocks with conditionally selected ancillas based on mid-circuit erasure checks, a key part of leakage-robust error correction schemes using neutral atoms. 
\end{abstract}

\maketitle
\onecolumngrid

\section{Introduction}

Recent advances in scaling quantum processors have led to breakthrough demonstrations of quantum error correction (QEC) and fault-tolerant computing in many hardware platforms~\cite{ryan-anderson_realization_2021, egan_fault-tolerant_2021, gupta_encoding_2024, ryan-anderson_high-fidelity_2024, bluvstein_logical_2024,paetznick_demonstration_2024,reichardt_logical_2024,google_quantum_ai_and_collaborators_quantum_2025}. At the same time, advances in the design of quantum error correcting codes and fault-tolerant circuits~\cite{bravyi_high-threshold_2024,zhou_algorithmic_2024,gidney_magic_2024} and quantum algorithms~\cite{gidney_efficient_2019,bauer_quantum_2020,caesura_faster_2025} have reduced the apparent computational cost of running large-scale algorithms for real-world applications. Nevertheless, there is a significant gap between the projected requirements and the capabilities of current hardware, which has motivated efforts to improve physical qubit performance and scale.

A complementary approach to reducing the overhead of fault-tolerant computing is to engineer qubits such that the inevitable errors are of a type that can be efficiently corrected. Qubits with biased Pauli noise can be used to raise the threshold and lower the overhead~\cite{aliferis_fault-tolerant_2008, bonilla_ataides_xzzx_2021}, motivating recent studies of bosonic qubits~\cite{puri_bias-preserving_2020,reglade_quantum_2024,putterman_hardware-efficient_2025}. Another approach is to engineer qubits with predominantly erasure errors~\cite{grassl_codes_1997,bennett_capacities_1997} by shaping the physical errors into detectable leakage outside of the computational subspace~\cite{campbell_certified_2020,wu_erasure_2022,kubica_erasure_2023}. This can lead to significantly higher thresholds and improved sub-threshold scaling~\cite{barrett_fault_2010,wu_erasure_2022,kubica_erasure_2023,sahay_high-threshold_2023}. A similar benefit can be achieved by shaping errors into qubit loss (i.e., atoms ejected from the trap), given the availability of loss-resolving measurements~\cite{baranes_leveraging_2025,yu_processing_2025}.

This has motivated significant experimental work to design physical qubits with an erasure error bias. To date, these demonstrations have relied on postselection, demonstrating improved memory, single- or two-qubit gate fidelity in the absence of detected erasures ~\cite{ma_high-fidelity_2023, scholl_erasure_2023, levine_demonstrating_2024, chou_superconducting_2024, shi_long-lived_2025, huang_logical_2026}  or qubit loss~\cite{radnaev_universal_2025, muniz_high-fidelity_2024}, or improved postselected performance of logical qubits encoded in error detecting codes when decoding with loss information~\cite{reichardt_logical_2024}. However, no experiment to date has demonstrated logical error suppression without postselection using erasure information.

In this work, we demonstrate QEC primitives using qubits encoded in the metastable $^3\text{P}_0$ state of \yybbns, which allows fast detection of certain errors as erasures with very little back-action onto qubits remaining in the computational space~\cite{wu_erasure_2022,ma_high-fidelity_2023}. The execution of programmable circuits is facilitated by a zone-based architecture~\cite{bluvstein_logical_2024}, and we demonstrate several unique capabilities of \yybb in this context. First, decoherence during moving is strongly suppressed by using the nuclear spin qubit and with appropriate control of the dipole trap polarization, leaving an error channel that is strongly biased towards loss. Second, single-photon Rydberg excitation enables two-qubit gates that are robust to intensity variations across the gate zone~\cite{jandura_optimizing_2023,fromonteil_protocols_2023}. We leverage these abilities to demonstrate logical qubit encoding in $[[4,2,2]]$ codes, and show improvement in the logical qubit decoding by using mid-circuit erasure information, with and without postselection. In addition, we demonstrate teleportation of logical qubits between distinct code blocks using conditionally selected ancillas based on mid-circuit erasure checks, which is a key building block for leakage- and loss-robust error correction protocols.
\\
\section{Results}

\begin{figure*}
    \centering
    \includegraphics[width=180 mm]{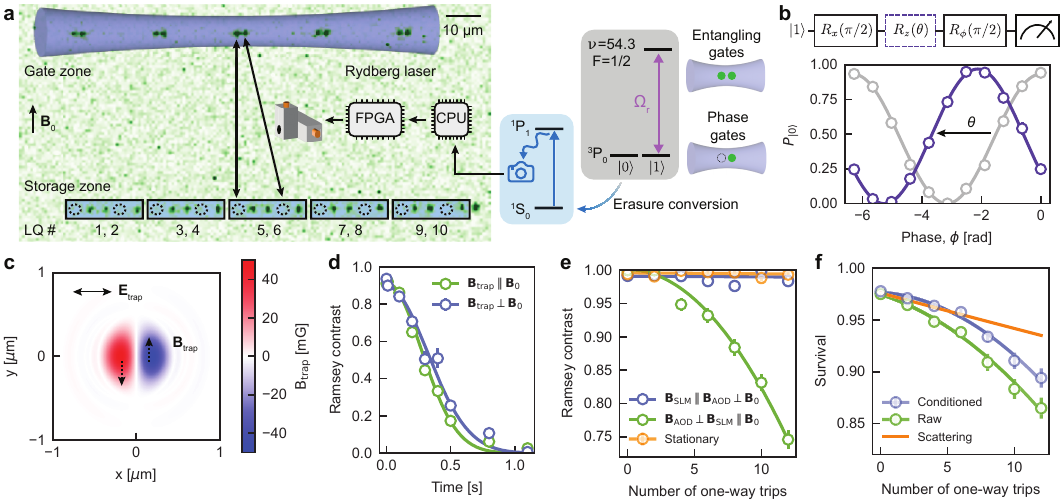}
    \caption{{\bf Logical qubit architecture and coherent transport with metastable $^{171}$Yb qubits.}
    (a) Schematic of storage and gate zones for implementing the [[4,2,2]] code. Qubits are encoded in the nuclear spin sublevels of the metastable $^3P_0$ manifold, and leakage errors to the $^1S_0$ ground state can be detected using fast imaging on the $^1P_1$ transition. Coherent atom transport between the two zones is implemented using AODs controlled by an FPGA, which allows the circuit execution to be adapted based on mid-circuit erasure information. A tightly focused Rydberg laser (302~nm  wavelength, 12~$\mu$m $1/e^2$ radius) defines the gate zone used to implement both one- and two-qubit gates. Two-qubit gates use resonant excitation from $\ket{1}$ to the Rydberg state $\ket{r}=\ket{6s\nu s, \nu=54.3, F=1/2, m_F=-1/2}$.
    (b) Selectively applied single-qubit gate $R_{z}(\theta)$ applied in the gate zone (purple), while the atoms in storage zone remain unaffected (gray).
    (c) The metastable qubit experiences a vector light shift near the tweezer focus that induces a synthetic magnetic field of approximately 50 mG, oriented in the focal plane, perpendicular to the tweezer polarization.
    (d) Coherence measured with a Ramsey sequence in stationary tweezers with the synthetic field perpendicular to the bias field ($T_2^* = 0.45(2)$ s) or parallel to it ($T_2^* = 0.39(1)$ s).
    (e) Coherence measured with a Ramsey sequence while transporting atoms between the storage and gate zones when the synthetic field of the SLM tweezers is parallel (green) or perpendicular (blue) to the external field, compared to an atom held in a stationary tweezer for the same duration (orange). A one-way trip includes handing off atoms from the SLM to AOD tweezers and back.
    (f) Atom loss from the metastable state while transporting between the storage and gate zones, before (green) and after (blue) conditioning on the absence of a mid-circuit erasure detection. The expected scattering contribution is shown in orange. 
    Approximately half of the loss is detectable over the first three round trips.
    Heating-induced loss lowers the erasure fraction to $\sim10\%$ by the fifth round trip. In all panels, the error bars show $\pm 1 \sigma$ statistical uncertainty.
}
    \label{fig:intro}
\end{figure*}

We implement programmable quantum circuits by transporting atoms between a storage zone and a gate zone (Fig.~\ref{fig:intro}a)~\cite{bluvstein_logical_2024,reichardt_logical_2024}. The static traps defining both the storage and gate zones are generated by a liquid crystal spatial light modulator (SLM), while the dynamic transport traps that govern the circuit execution are generated by a pair of orthogonally oriented acousto-optic deflectors (AODs). The AODs are driven by piecewise-polynomial waveforms generated in real-time using a field-programmable gate array (FPGA)~\cite{stefanazzi_qick_2022}, configured to receive mid-circuit erasure check information from a fast EMCCD camera through an intermediate computer. The gate zone is illuminated by a focused 302 nm laser beam used to implement entangling Rydberg gates via the $\ket{\nu = 54.3}$ state (Fig.~\ref{fig:intro}a; $P = 20$\ mW, $\Omega_r = 2\pi \times 2.5$\ MHz)~\cite{peper_spectroscopy_2025}. For convenience, we also use the Rydberg transition to implement selective single-qubit $Z(\theta)$ rotations (Fig.~\ref{fig:intro}b). Complemented by a global radio-frequency drive, these operations provide full single-qubit control. Since quantum circuits require repeated trap switch-offs for light-shift–sensitive operations, we use periodic trap modulation to avoid the heating from irregular switching and enable deeper circuit execution (Methods).

Minimizing qubit errors during transport is critical in a reconfigurable processor. Implementations with alkali atoms make use of rapid dynamical decoupling interleaved with the atomic motion to suppress errors from differential light shifts~\cite{beugnon_two-dimensional_2007,bluvstein_quantum_2022,manetsch_tweezer_2024}. Rapid dynamical decoupling is challenging with the nuclear spin qubit in Yb, however, seconds-scale coherence times can be achieved without it in both the \ssz and \tpz states~\cite{ma_universal_2022,jenkins_ytterbium_2022,barnes_assembly_2022,ma_high-fidelity_2023,lis_midcircuit_2023}. Compared to the \ssz ground state, the \tpz state has a much larger vector light shift from elliptical polarization~\cite{porsev_possibility_2004}. The tweezer light is elliptically polarized near the focus from non-paraxial effects~\cite{thompson_coherence_2013}, which generates a synthetic field varying across the tweezer with maximal magnitude $|\vec{B}_{\text{trap}}| = 50$\,mG at a typical trap depth (Fig.~\ref{fig:intro}{\color{blue}c}). This can split the qubit levels by $\Delta \nu = 57$ Hz if the external bias field $\vec{B}_0$ is parallel to $\vec{B}_\text{trap}$; the effect can be reduced by applying the bias field in the perpendicular direction (here, $|\vec{B}_0| = 5$\,G suppresses $\Delta \nu$ to $0.3$ Hz). Either configuration is sufficient to achieve long coherence times in stationary traps, where the atomic motion is centered on the focus (Fig.~\ref{fig:intro}d). However, any small misalignment between the two traps displaces the potential minimum from the beam focus during the handoff. As a result, even with careful alignment, we still observe strong dephasing when handing off atoms between traps when either trap is configured with its synthetic field parallel to $\vec{B}_0$ (green curve in Fig.~\ref{fig:intro}e, as $\vec{B}_\mathrm{SLM}\parallel \vec{B}_0$).

Alternatively, with the optimal polarization configuration ($\vec{B}_{\text{SLM}}\parallel\vec{B}_\text{AOD}\perp\vec{B}_0 $), we observe a combined dephasing and bit flip error probability of $< 6 \times 10^{-4}$ (Fig.~\ref{fig:intro}e) per one-way trip from the storage zone to the gate zone, measured by the Ramsey fringe contrast of surviving atoms. A one-way trip lasts approximately 2.5 ms and includes handing off atoms between the stationary and moving tweezers on each side, without any dynamical decoupling. The error per trip is determined by repeating up to 12 trips. Photon scattering (Raman scattering and photoionization~\cite{ma_high-fidelity_2023}) and heating contribute additional errors in the form of leakage from \tpz and atom loss. Over the first round trip (comprising four trap handoffs and two shuttling steps), the average loss probability (including both leakage and atom loss) from \tpz is 1.1(4)\% (Fig.~\ref{fig:intro}f), which increases in subsequent trips because the atom is heated. Before the onset of heating, roughly half of this error is detectable mid-circuit as an erasure error.

Next, we turn to two-qubit gates. A common challenge to laser-driven gates is laser intensity variations, which can arise from laser noise, pointing instability or inhomogeneity across a gate zone, and our day-to-day gate fidelity is limited by pointing stability of the Rydberg gate laser. To circumvent this challenge, we implement a variant of the usual symmetric CZ gate~\cite{levine_parallel_2019,jandura_time-optimal_2022} that is robust against quasi-static laser intensity variations, called the amplitude-robust (AR) gate (Fig.~\ref{fig:arcz}a -- c)~\cite{jandura_optimizing_2023,fromonteil_protocols_2023}. For a fractional intensity error $\Delta I /I_0$, the AR gate error scales as $\epsilon \propto (\Delta  I/I_0)^4$ instead of $(\Delta  I/I_0)^2$ for conventional gates, at the expense of spending 60\% more time in the Rydberg state. The AR gate requires that the AC Stark shift of the Rydberg transition $\Delta_{LS} = \chi \Omega_r^2$ is smaller than the Rabi frequency~\cite{jandura_optimizing_2023}, which is not satisfied for state-of-the-art two-photon gates~\cite{evered_high-fidelity_2023} but is readily satisfied for single-photon gates in \yybb (Fig.~\ref{fig:arcz}d). The AR gate fidelity at the typical laser power is $\mathcal{F}_{\text{AR}}=0.984(1)$, which is 1.45 times more error than the time-optimal (TO) gate fidelity of $\mathcal{F}_{\text{TO}}=0.989(1)$ under the same conditions (Fig.~\ref{fig:arcz}e). Remarkably, the AR gate fidelity is nearly unchanged when deliberately reducing the gate laser intensity by a significant amount without recalibration (Fig.~\ref{fig:arcz}f). This robustness allows us to extend the recalibration interval from hours to days, and all subsequent experiments use the AR CZ gate. Moreover, 38(6)\% of the AR gate error is detected mid-circuit as erasure errors (Fig.~\ref{fig:arcz}e). All CZ-gate benchmarking is performed using the same randomized-benchmarking-type protocol introduced in Ref.~\cite{peper_spectroscopy_2025}, interleaving CZ gates with random global single-qubit Clifford gates (Extended Data Fig.~\ref{fig:grape}e).

We now proceed to study erasure conversion in the context of the $[[4,2,2]]$ error detecting code. As a $d=2$ code it is typically used for error detection~\cite{linke_fault-tolerant_2017,andersen_repeated_2020,gupta_encoding_2024}, but it can also correct a single erasure error~\cite{grassl_codes_1997}. The codespace is stabilized by $S_Z = Z^{\otimes 4}$ and $S_X = X^{\otimes 4}$, and the logical operators are given by $X_L^{(1)} = X_1 X_3$ and $X_L^{(2)} = X_1 X_2$, and $Z_L^{(1)} = Z_1 Z_2$ and $Z_L^{(2)} = Z_1 Z_3$ (Fig.~\ref{fig:code422}a)~\cite{grassl_codes_1997}. The encoding circuit in Fig.~\ref{fig:code422}b is used to prepare the $\ket{00}_L$ or $\ket{++}_L$ state, with an ancilla flag that can be used to detect certain high weight Pauli and leakage errors (Methods). After a variable hold time $t$, we perform a mid-circuit erasure check to detect any erasure errors during the preparation circuit or idle time, then make a final transversal measurement of all qubits and decode.

When using the code for error detection without erasure information, we achieve a logical qubit state preparation fidelity averaged over $\ket{00}_L$ and $\ket{++}_L$ of 0.981(2), which increases to 0.990(1) when also post-selecting on the flag qubit (Fig.~\ref{fig:code422}c). Additionally post-selecting on the mid-circuit erasure information improves the initialization to 0.995(1). The post-selected error rate grows quadratically with time during the hold period in all cases, but the rate is 3.6(1) times slower when including the erasure information. The data is in good agreement with a simulation based on independently measured errors (Methods).

Importantly, mid-circuit erasure information can be used to improve the unconditional decoding of the $d=2$ code, increasing the probability to decode the correct logical state without post-selection (Fig.~\ref{fig:code422}e). The state preparation fidelity increases from 0.874(4) to 0.902(3) when erasure information is included, and the decay rate with increasing hold time is reduced by a factor of 1.9(4). 
Notably, the resulting decay rate with erasure information is comparable to that of a single physical qubit under identical conditions, even though the logical qubit consists of multiple physical qubits that accumulate errors more rapidly; however, initialization errors still limit the absolute fidelity.
For this code, the decoding rule with erasure information is simple: if the parity is even, no correction is applied; if the parity is odd and exactly one erasure is present, we flip the erased qubit.

\begin{figure*}
    \centering
    \includegraphics[width=180 mm]{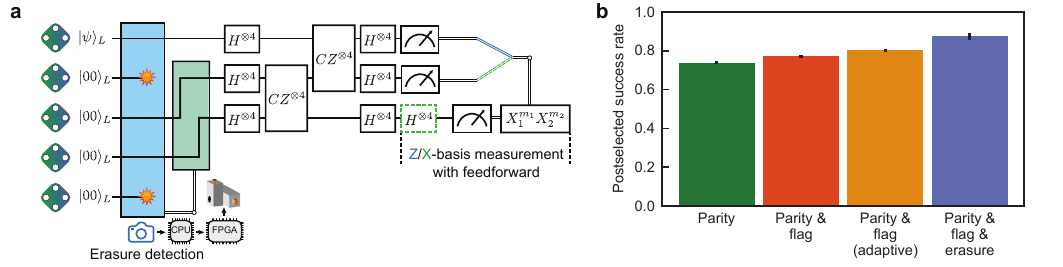}
    \caption{{\bf Logical qubit teleportation with adaptive ancilla selection.}
    (a) After state preparation, a mid-circuit erasure check is used to select two ancillas without erasure errors to prepare a logical Bell state. The input logical state is teleported to the second ancilla block and decoded in the Z or X basis.
    (b) Teleportation success probability when post-selecting the final parity of all logical qubits, parity and flag, or further the absence of detected erasure errors, averaged over teleporting $|00\rangle_L$ and $|++\rangle_L$. Logical ancilla blocks used to form the Bell pair are chosen at random in all cases except for the yellow bar, where ancillas are selected adaptively.
    The teleportation circuit with random ancilla-block selection is run (5458, 6024) times for the $\ket{00}_L$ and $\ket{\text{++}}_L$ inputs, respectively; with adaptive ancilla selection, the corresponding sample numbers are (5348, 6006). Error bars in (b) indicate statistical uncertainty ($\pm 1\sigma$).
    }
    \label{fig:knill}
\end{figure*}

Finally,  we demonstrate teleportation of logical qubits~\cite{chou_deterministic_2018,erhard_entangling_2021,postler_demonstration_2022,ryan-anderson_high-fidelity_2024,bluvstein_logical_2024}, an important QEC primitive for error correction in the presence of atom loss. In this experiment, logical Bell pairs are prepared between two ancilla code blocks, and the logical data qubits are teleported using a circuit equivalent to Knill's teleportation circuit~\cite{knill_quantum_2005} (Fig.~\ref{fig:knill}a). As a start, we first randomly select two out of four available code blocks as ancillas (Fig.~\ref{fig:knill}b). When post-selecting on trivial syndrome values in all logical qubits and the absence of flags, the teleportation fidelity is 0.771(9), averaged over the $\ket{00}_L$ and $\ket{++}_L$ input states. It rises to 0.87(2) when also post-selecting on the absence of erasures.

When erasures are detected in the preparation of the ancilla qubits, the affected logical blocks can be discarded instead of attempting to repair the errors. We demonstrate this using adaptive control over the atom trajectories to select two out of the four ancilla blocks without a detected erasure. We observe a modest improvement in the teleportation fidelity, from 0.771(9) to 0.802(8), with post-selection on both trivial syndrome values and the flag qubit. 
This limited gain results from the fact that the distance-2 demonstration is primarily limited by Pauli errors instead of erasures.
Although the improvement is therefore modest in the present system, postselected resource states are expected to yield significantly higher thresholds when erasure errors are dominant~\cite{bartolucci_fusion-based_2023, sahay_high-threshold_2023, paesani_high-threshold_2023}.
\\
\section{Discussion}

In this work, we have demonstrated a zoned architecture for metastable $^{171}$Yb qubits leveraging mid-circuit erasure error detection, low-decoherence atom transport and robust entangling gates. We implemented several QEC primitives with erasure conversion, including error correction during decoding using an error-detecting code and logical state teleportation between code blocks. Overall, our results show a clear benefit of erasure conversion in error-correction circuits: even with a small $d=2$ code, we observe improved logical-qubit performance without any postselection, unambiguously demonstrating the advantages of erasure conversion. This work also provides a first demonstration of adaptive circuit execution conditioned on erasure information.

\begin{figure}
    \centering
    \includegraphics[width=90 mm]{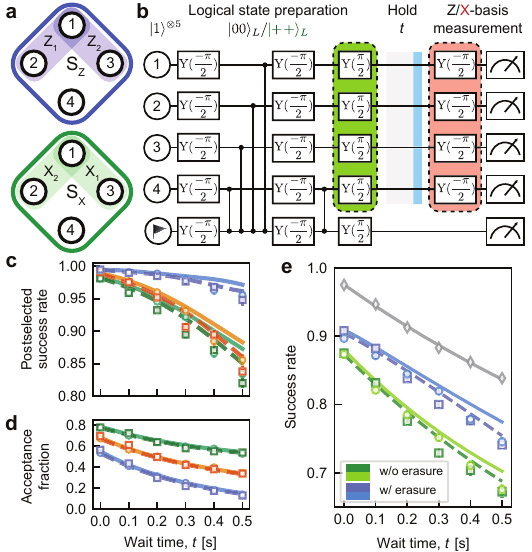}
    \caption{{\bf$[[4,2,2]]$ code implementation.}
    (a) Stabilizers and logical operators of the $[[4, 2, 2]]$ code. 
    (b) Encoding circuit for $\ket{00}_L$ and $\ket{++}_L$. The latter is obtained by applying transversal $R_y(\pi/2)$ gates to the four data qubits (green box). After a waiting period $t$, an erasure check (blue) is followed by transversal measurements on all qubits and decoding of the logical state in the $Z$ or the $X$ basis.
    A flag qubit is used to catch certain high-weight Pauli and leakage errors.
    (c) Post-selected state preparation and memory fidelity after preparing $\ket{00}_L$ (round markers) or $\ket{++}_L$ (square markers), shown alongside simulation results (solid and dashed curves). Green data corresponds to post-selection on the measured parity (i.e., stabilizer value) alone; orange includes both parity and flag qubit detection in the bright state ($\ket{0}$); blue further conditions on the absence of detected erasure errors. The success probability describes the probability to decode a single logical qubit correctly. 
    (d) Acceptance fractions corresponding to the different post-selection choices used in (c).
    (e) State preparation and memory fidelities without post-selection for $\ket{00}_L$ (circles) and $\ket{\text{++}}_L$ (squares), shown alongside simulation results. The green data corresponds to decoding only using the qubit measurement data, while the blue data incorporates the erasure information. The solid (dashed) curves overlapping with the data points with matched colors represent the simulation results for $\ket{00}_L$ ($\ket{\text{++}}_L$) states.
    The gray diamonds show the preparation and memory fidelity of a single physical qubit.
    For wait times from 0 to 0.5~s, the total sample counts are (2084, 2283), (2088, 2285), (2170, 2289), (2161, 2165), (2059, 2184), and (1968, 2299), respectively, where each pair corresponds to the $\ket{00}_L$ and $\ket{\text{++}}_L$ preparations. Error bars in (c–e) indicate statistical uncertainty ($\pm 1\sigma$) and are sometimes smaller than the markers. 
    }
    \label{fig:code422}
\end{figure}

Despite these results, further improvements in logical-circuit performance will require identifying the dominant error mechanisms in the current system. Our numerical model indicates that transport errors dominate the present circuit implementation. For example, in Fig.~\ref{fig:code422}e, the model suggests that transport errors account for 66\% of the logical error rate at $t=0$. We emphasize that this limitation is specific to the present circuit implementation, which lacks an alternative method for individual addressing and, due to the 1D gate zone, offers only limited parallelism. As a result, each atom requires four moves between the storage and gate zones, and executing these moves sequentially introduces a substantial temporal overhead (29 ms).

Beyond technical improvements, practical fault-tolerant quantum computation ultimately requires larger-distance codes that are also robust against Pauli errors. One suitable example is the surface code, where numerical studies suggest that erasure conversion can significantly improve its threshold and effective error distance \cite{wu_erasure_2022, sahay_high-threshold_2023}. Even with the current erasure fraction of our CZ gates (about 38\%), erasure conversion is predicted to raise the surface-code threshold from 0.94\% to 1.45\%, and thresholds over 10\% have been projected for very high erasure fractions~\cite{sahay_high-threshold_2023}. The high parallelism of logical operations in the surface code will also help reduce errors from atom transport. For example, in a 2D gate-zone architecture incorporating separate trap arrays for data qubits and ancillas, a single surface-code syndrome-extraction cycle could be executed with no handoffs and only four short-distance moves~\cite{bluvstein_quantum_2022}. This approach is expected to reduce heating and enable faster circuit execution.

\begin{figure*}
    \centering
    \includegraphics[width=130 mm]{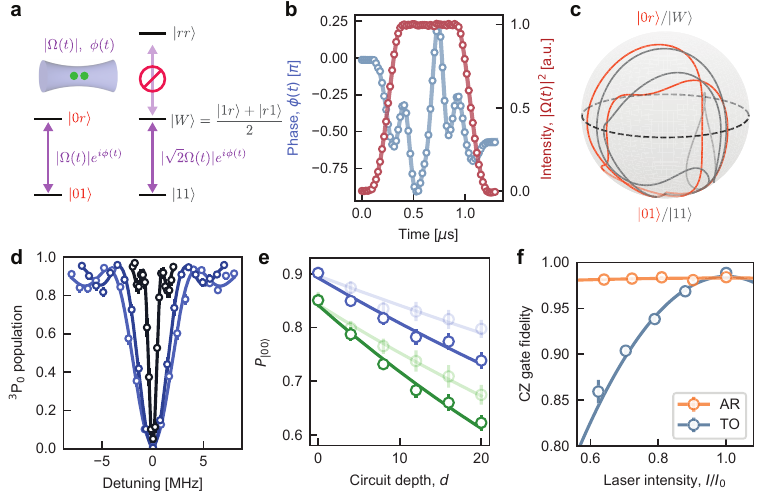}
    \caption{{\bf Amplitude-robust CZ gates.}
    (a) The AR gate is a symmetric CZ gate that uses a phase-modulated drive to implement simultaneous closed trajectories when starting from $\ket{01}$ (or $\ket{10}$) and $\ket{11}$, leveraging the $\sqrt{2}$ enhancement of the Rabi frequency from the blockaded excitation to $\ket{rr}$ in the latter case to implement an entangling gate~\cite{levine_parallel_2019,jandura_time-optimal_2022}.
    (b) AR gate intensity (red) and phase (blue) profiles, measured with a heterodyne interferometer~\cite{ma_high-fidelity_2023}. The ideal waveforms are shown with solid lines.
    (c) Bloch sphere trajectories during the AR gate in the $\{\ket{01},\,\ket{0r}\}$ subspace (orange) and $\{\ket{11},\ket{W}\}$ subspace (grey).
    (d) Measurement of the $\ket{1}$ -- $\ket{r}$ transition at different laser powers ranging from $\Omega_r/(2\pi) = 0.5$ MHz to $2.5$ MHz. No shift is observed, bounding the light shift coefficient to $\chi < 3$\,kHz/MHz$^2$.
    (e) Randomized circuit benchmarking~\cite{evered_high-fidelity_2023,ma_high-fidelity_2023,tsai_benchmarking_2025} of AR gates with mid-circuit erasure detection.
    The extracted error rate of the AR CZ gate is 
    $\epsilon_{\rm{AR}} = 0.016(1)$ (solid blue), which is reduced to $\epsilon_{c, {\rm{AR}}} = 0.010(1)$ (solid green) after conditioning on not detecting an atom in the ground state before the end of the circuit. Under the same conditions, the TO gate errors (light curves) are $\epsilon_{\rm{TO}} = 0.011(1)$ and $\epsilon_{c,\rm{TO}} = 0.006(1)$.
    (f) CZ gate fidelity as a function of the actual intensity $I$ relative to the nominal calibration point $I_0$. The solid curves show phenomenological fits to $[(I_0 - I)/I_0]^2=(\Delta I/I_0)^2$ and $(\Delta I/I_0)^4$ for the TO and AR data, respectively. In panel (d-f), the error bars show the statistical uncertainty ($\pm1 \sigma$).}
    \label{fig:arcz}
\end{figure*}

In an operational fault-tolerant quantum computer, repeated error correction and deep logical circuits will require replacing the affected atoms, as well as atoms that were lost without being detected, using leakage reduction units~\cite{aliferis_fault-tolerant_2006}. The most natural approach for neutral atoms is to teleport logical information to freshly prepared physical qubit arrays, as considered by Knill~\cite{knill_quantum_2005} and in measurement-based quantum computing implementations~\cite{bartolucci_fusion-based_2023,sahay_high-threshold_2023}. It has recently been proposed that loss-resolving measurements~\cite{lis_midcircuit_2023, norcia_midcircuit_2023, chow_circuit-based_2024,li_fast_2025} can provide erasure-like information about qubit loss~\cite{gu_optimizing_2024,chang_surface_2024, perrin_quantum_2025, yu_processing_2025, baranes_leveraging_2025}. This can result in improved erasure detection performance: for example, we have measured that 75\% of the TO gate errors in Fig. \ref{fig:arcz}e result in atom loss, and virtually all of the transport errors are loss. To ensure a supply of new physical qubits, a key step for the future of this platform will be fast mid-circuit atom replacement to allow loss errors to be replenished~\cite{singh_dual-element_2022,muniz_repeated_2025,li_fast_2025}.

Lastly, we note that while completing this work, we became aware of complementary work on \yybbns, demonstrating metastable qubit gates with atom loss detection~\cite{senoo_high-fidelity_2025}, and work demonstrating improved logical qubit performance using a loss-aware decoder~\cite{bluvstein_fault-tolerant_2026}.

\vspace{10pt}

\textbf{Acknowledgements} We acknowledge the QICK team at Fermilab (Gustavo Cancelo, Leandro Stefanazzi) for developing the FPGA tweezer controller, and Sara Sussman for the help with the initial implementation. We also thank Sven Jandura, Andrew Ludlow, Kyle Beloy and Adam Kaufman for helpful conversations. This work was supported by the Army Research Office (W911NF-18-10215, W911NF-24-10358), DARPA MeasQuIT (HR00112490363) and ONISQ (W911NF-20-10021), the Office of Naval Research (N00014-23-1-2621), and the National Science Foundation through the CAREER program (PHY-2047620) and the Center for Robust Quantum Simulation (OMA-2120757).

\vspace{10pt}

\textbf{Author Contributions} B.Z., G.L., G.B., S.P.H., P.P., and S.M. built the experimental setup, performed the measurements and analyzed the data. S.H. and S.P. contributed to circuit constructions, data analysis, and stabilizer circuit simulations. All authors discussed the results. B.Z., G.L., G.B., and J.D.T. wrote the manuscript with input from all authors. The work was supervised by S.P. and J.D.T.

\vspace{10pt}

\textbf{Competing interests} S.P.H and J.D.T are co-founders and shareholders of Logiqal, Inc. J.D.T., B.Z., G.L., G.B., P.P., S.P.H., and S.M. have filed provisional patents directly arising from the results detailed in this manuscript. The remaining authors declare no competing interests.

\clearpage

\clearpage


\onecolumngrid

\clearpage
\renewcommand{\theequation}{S\arabic{equation}}
\renewcommand{\figurename}{Extended Data Fig.}
\renewcommand{\tablename}{Extended Data Table}

\setcounter{figure}{0}
\setcounter{equation}{0}

\section*{Methods}

\subsection*{Experimental implementation}
\label{sec:exp_implement}

Our apparatus has a static and a dynamic optical tweezer array. The static path is generated by a spatial light modulator (SLM, Hamamatsu model X13138-04WR) and has 88 traps: 3 $\times$ 26 traps for storage and loading, and 10 in the gate zone. The dynamic path is generated by a pair of orthogonal acousto-optic deflectors (AODs). Both optical paths are derived from the same laser source (Coherent Genesis CX-488) at a wavelength of $\lambda = 487$\,nm with opposite AOM frequency shifts to avoid interference. For most experiments, the two paths have orthogonal linear polarizations and are combined via a polarizing beamsplitter.

The physical qubits are encoded in the metastable $6s6p \, ^3\text{P}_0$ states of neutral \yybbns, where $\ket{0} \equiv \ket{F=1/2, m_F=-1/2}$ and $\ket{1} \equiv \ket{F=1/2, m_F=1/2}$. The metastable state is  initialized and measured using optical pumping, following the procedure in Ref.~\protect\citesupp{ma_high-fidelity_2023}. We have improved the efficiency of the pumping process by applying the pumping while modulating the optical dipole trap (see \textit{Periodic trap modulation}), which eliminates light shifts during the pumping processes. With this approach, we observe a round-trip pumping loss of 0.6\%. We believe the majority of the losses arise in the depumping step.

The lifetime of the metastable state is measured to be $1.64(3)$ s, under a trap power of approximately $1.5$ mW per tweezer and is primarily limited by the Raman scattering and photoionization from the trap light. The spin-flip time $T_1$ exceeds $13$ s, and the coherence time is measured to be $T_2^* = 0.39(1)$ s (with $T_2 = 6(2)$ s).

Global single-qubit gates are driven by a radio-frequency (RF) magnetic field. Under our applied bias magnetic field of $B_0 = 5$\,G, the Larmor frequency between $\ket{0}$ and $\ket{1}$ is $\omega_L = 2\pi\times 5.7$\,kHz. To operate within the rotating-wave approximation regime, we employ a relatively low Rabi frequency of $\Omega_\text{SQ} \sim 2\pi\times 200$\,Hz, but this restriction is not fundamental~\protect\citesupp{rower_suppressing_2024}. We measure a fidelity of $\mathcal{F}_{\text{1Q}} = 0.9990(1)$ for RF single-qubit gates using randomized benchmarking.

A focused 302 nm laser ($1/e^2$ radius $w_0 = 12\,\mu\text{m}$) defines the gate zone for performing two-qubit gates~\protect\citesupp{bluvstein_quantum_2022, bluvstein_logical_2024}. For convenience, we implement selective single-qubit gates using the same beam, performing $Z$ rotations via the Rydberg state. Selective projective measurements are also implemented in this zone by blowing out population in the $\ket{1}$ state through excitation to the Rydberg state and subsequent autoionization~\protect\citesupp{ma_universal_2022, ma_high-fidelity_2023}, though the population remaining in $\ket{0}$ is not imaged until the end of the circuit.

Following Ref.~\protect\citesupp{ma_high-fidelity_2023}, two different cycling transitions are used to measure information about the qubits. For initialization and final measurement, we use the intercombination line \ssz $\rightarrow$ $^3\text{P}_1$ at 556 nm, achieving a fidelity and survival probability of 0.995 with a 15 ms exposure time. For the mid-circuit erasure detection, we use the \ssz $\rightarrow$ $^1\text{P}_1$ transition at 399 nm, achieving a detection fidelity of $\approx 0.99$  in 20 $\mu$s at the expense of heating the atom out of the optical dipole trap. Because of the short exposure time and large detuning, the erasure detection has negligible back action on qubits remaining in the metastable state.

\subsection*{Real-time AOD waveform generation}
\label{sec:rfsoc}

\begin{figure*}
    \includegraphics[width=120mm]{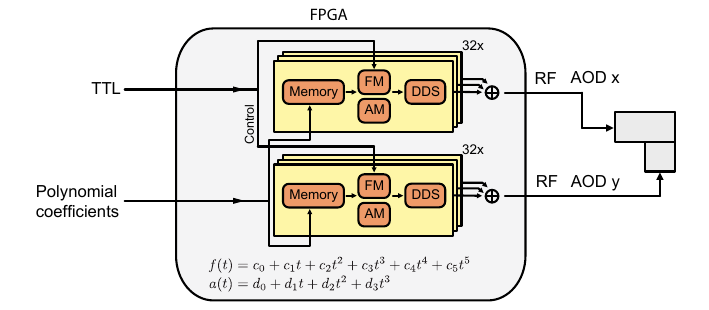}
    \caption{{\bf Real-time waveform generation.} The AOD waveforms are generated on a Xilinx RFSoC ZCU216 evaluation board. The trajectories are expressed as piecewise polynomial coefficients in frequency and amplitude. The coefficients are transmitted to the FPGA over ethernet, and the execution of each segment is triggered by a TTL line. Two output channels drive orthogonally oriented AODs, each supporting up to 32 DDS tones.}
    \label{fig:rfsoc}
\end{figure*}

The RF for the acousto-optic deflector is generated by a ZCU216 RFSoC FPGA evaluation board programmed with a variant of the QICK firmware \protect\citesupp{stefanazzi_qick_2022}. The X/Y channel outputs are generated by summing 32 independent DDS generators (Extended Data Fig.~\ref{fig:rfsoc}). Each DDS outputs an independent RF tone and can be programmed to perform a frequency chirp described by a piecewise fifth-order polynomial (allowing trajectories such as min-jerk~\protect\citesupp{flash_coordination_1985} to be specified as a single segment), along with a third-order polynomial for the amplitude. This allows waveform information to be transmitted much more efficiently than a raw waveform~\protect\citesupp{young_half-minute-scale_2020}.
Sequence synchronization is achieved using a real-time processor on the FPGA, which receives streaming commands from a PC with a frame grabber that is used to process the camera images. We currently operate with a 25 ms latency between the end of the image exposure time and the beginning of the next step of waveform generation. Most of this delay is to accommodate worst-case latency depending on other processes running on the PC. The FPGA-generated waveforms are compatible with future developments to lower this latency, or to process images directly on the FPGA.

\subsection*{Atom loss and heating during transport}
\label{sec:move_loss}

During mid-circuit movement, atom transport can introduce heating or even result in atom loss. In our investigation, we identify a few contributing factors in our system.

\begin{figure*}
    \includegraphics[width=180mm]{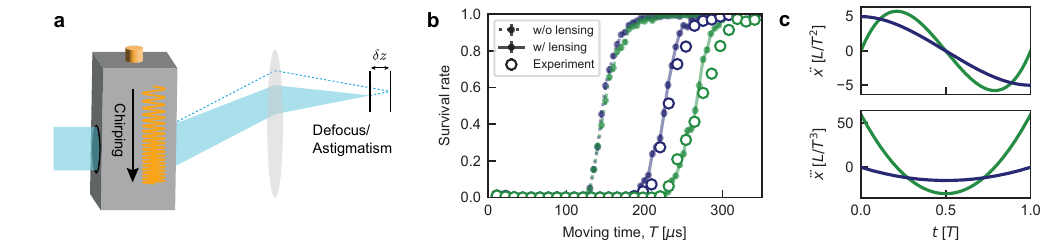}
    \caption{{\bf AOD lensing effect.} (a) Due to the finite shear mode acoustic speed in the AOD crystal and a large optical aperture, a non-negligible lensing effect occurs when chirping the AOD driving frequency, causing defocus and astigmatism.
    (b) Experimentally measured survival probability (circles) after a one-way trip between the storage and gate zones while changing moving time $T$, with the zero-jerk trajectory (blue) and minimum-jerk trajectory (green). Numerical simulation with (solid line) and without (dash line) considering the lensing effect are also shown. The simulation matches with experiment only if the lensing effect is included.
    (c) Acceleration and jerk profiles of the zero-jerk trajectory (blue) and minimum-jerk trajectory (green). }
    \label{fig:lensing}
\end{figure*}

Intuitively, transporting atoms too quickly leads to increased heating and atom loss. This behavior is consistent with our experimental observations (Extended Data Fig.~\ref{fig:lensing}b), where we hold atoms in AOD traps and perform a single transport from the storage to the gate zone. To identify the primary limitation in this experiment, we develop a numerical model based on the Monte Carlo method to simulate the classical dynamics of atoms during transport. Interestingly, we find that the dominant limitation is not radial heating from the trap, but rather the ``acoustic lensing effect'' (Extended Data Fig.~\ref{fig:lensing}a~\protect\citesupp{dickson_optical_1972, manetsch_tweezer_2024}). This effect arises from the finite response time of the AOD: as the driving frequency is swept, the AOD acts like a cylindrical lens, introducing astigmatism in the optical trap.

The focal shift $\delta z(t)$ induced at the optical trap can be expressed as
\begin{equation}\label{eq:lensing}
    \frac{\delta z(t)}{z_R} = \frac{w_\text{aod}}{v_s} \frac{\dot{x}(t)}{w} = \tau_\text{aod} \frac{\dot{x}(t)}{w},
\end{equation}
where $z_R$ and $w$ are the Rayleigh range and beam waist of the optical tweezer in the atomic plane, respectively, while $w_\text{aod}$ and $v_s$ denote the AOD aperture (radius) and the speed of sound in the AOD material, and  $\tau_\text{aod} = w_{aod}/v_s$ is the AOD output rise time. Cylindrical lensing begins to be significant when the transport speed exceeds one beam waist per $\tau_\text{aod}$, and is therefore more pronounced for shorter trap wavelengths when moving a fixed distance.

In the experiment, we study this acoustic lensing effect by varying the transport trajectory between the storage and gate zones. As described in \textit{Real-time AOD waveform generation}, we parameterize the trajectories using a fifth-order polynomial. With additional symmetry constraints, the trajectory takes the form
\begin{equation}\label{eq:traj}
    x(t) = L\left[(15-8\gamma)\left(\frac{t}{T}\right)^2+(-50+32\gamma)\left(\frac{t}{T}\right)^3+(60-40\gamma) \left(\frac{t}{T}\right)^4+(-24+16\gamma)\left(\frac{t}{T}\right)^5\right], \quad \left(0 \leq \frac{t}{T} \leq 1\right),
\end{equation}
where $T$ and $L$ are the duration and distance of transport, and $\gamma$ corresponds to the velocity at $t=T/2$ (in unit of $L/T$) and can be freely tuned. We compare two specific trajectories: $\gamma = 1.5625$ (we call it \textit{zero-jerk} trajectory since the jerk vanishes at both the start and end points, see Extended Data Fig.~\ref{fig:lensing}c) and $\gamma = 1.875$ (the minimum-jerk trajectory~\protect\citesupp{flash_coordination_1985}). In both cases, experimental results match the numerical model without any free fitting parameters only when the acoustic lensing effect is included (Extended Data Fig.~\ref{fig:lensing}b). Furthermore, the model predicts that with lensing taken into account, the zero-jerk trajectory outperforms the minimum-jerk one, which we verify experimentally. While in the absence of acoustic lensing, both trajectories perform similarly according to the simulation. We use the zero-jerk trajectory for the circuits in the rest of this work.

\begin{figure*}
    \includegraphics[width=180mm]{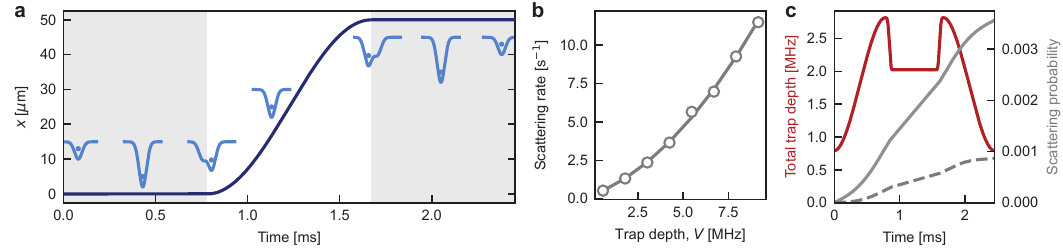}
    \caption{{\bf Scattering during atom transport.} (a) Schematic and trajectory of a one-way transport sequence: two hand-off phases between the static and moving optical tweezers (grey shading) bracket a moving phase (white).
    (b) Measured metastable state scattering rate versus trap depth, together with a quadratic fit~\cite{ma_high-fidelity_2023}.
    (c) Time trace of the total trap power experienced by the atom during the full trajectory (red); using the model in (b) we infer the cumulative probability of photo-ionization (grey dashed) and the total number of scattered photons (grey solid).}
    \label{fig:move_exp}
\end{figure*}

A one-way transport between the gate and storage zone in this work consists of three phases: (1) ramping up the AOD trap depth to handoff qubits from SLM traps to AOD traps, (2) moving, and (3) ramping down the AOD trap depth to handoff qubits back from AOD traps to SLM traps, as shown in Extended Data Fig.~\ref{fig:move_exp}a. To mitigate heating accumulated over multiple rounds of transport required by the circuit experiments, we use a longer moving time of 0.89 ms and a ramp time of 0.78 ms.

Another challenge for atom transport is scattering from the trap light. The \tpz metastable state qubits in particular suffer from both off-resonant scattering to other electronic states (such as the ground state) and photoionization \protect\citesupp{ma_high-fidelity_2023}. Increasing the AOD trap depth helps minimize heating and enable faster moving (Extended Data Fig.~\ref{fig:move_exp}a), at the cost of higher scattering rates (Extended Data Fig.~\ref{fig:move_exp}b and c). For the current move parameters, the expected scattering probability is about 0.3\% per one-way trip, which is roughly divided into 0.1\% per handoff and 0.1\% for the move, comparable to the total error rate for these steps quoted in other works~\protect\citesupp{bluvstein_logical_2024,reichardt_logical_2024}. The scattering probability depends strongly on the tweezer wavelength, and photoionization in particular vanishes for wavelengths longer than 604 nm.

\subsection*{Decoherence during trap handoff}
\label{sec:move_decohere}

One of the key advantages of using alkaline-earth(-like) atoms in optical tweezer arrays is the ability to encode qubits in nuclear spin states. In both the ground state (\( ^1\text{S}_0 \)) and the metastable clock state (\( ^3\text{P}_0 \)), the two valence electrons form a closed-shell configuration with total electronic angular momentum \( J = 0 \). Consequently, the hyperfine Zeeman sublevels \( \ket{m_F} \) are predominantly nuclear in character. Unlike hyperfine qubits in alkali atoms, where coherence is often limited by differential scalar light shifts between \( \ket{0} \) and \( \ket{1} \)~\protect\citesupp{kuhr_analysis_2005}, nuclear spin qubits in these \( J=0 \) states typically exhibit excellent coherence, with \( T_2^* \) timescales on the order of seconds, as demonstrated in this work and Ref.~\protect\citesupp{ma_universal_2022, jenkins_ytterbium_2022,ma_high-fidelity_2023,lis_midcircuit_2023}. The magnitude of differential light shifts is commonly characterized by the dimensionless parameter \( \eta = \delta E / V \), where \( \delta E \) is the energy difference between the two qubit states and \( V \) is the average (scalar) trap depth.

Although our system is free from differential scalar light shifts, they are still susceptible to vector light shifts. While the shift of the \ssz ground state is very small, the \( \ket{^3\text{P}_0, m_F = \pm 1/2} \) experience a much larger shift because of hyperfine mixing with nearby excited levels~\protect\citesupp{porsev_possibility_2004}. For our trapping wavelength, the vector light shift induced by fully circularly polarized light is estimated to be \( \eta_\mathrm{vls} \approx 3 \times 10^{-4} \)~\protect\citesupp{beloy_private_2025}.

This vector light shift introduces two primary challenges. First, any residual ellipticity in the trapping light, caused by imperfect polarization control or birefringence, can induce unwanted energy shifts. Second, even with linearly polarized input light, the high numerical aperture (NA) of the tweezer objective generates longitudinal polarization components near the focus, resulting in an overall elliptic polarization. Through the vector light shift, this generates a position-dependent synthetic magnetic field~\protect\citesupp{thompson_coherence_2013}:
\begin{equation}
    \vec{B}_\mathrm{trap}(r) = \frac{\eta_\text{vls} V}{g_m} \frac{\text{Im}[\vec E^*(r)\times \vec E(r)]}{|E_0|^2},
\end{equation}
where \( g_m = h \times 1.14 \) kHz/G characterizes the hyperfine splitting of the two metastable states, and \( \vec{E}(r) \) is the oscillating electric field normalized such that \( |\vec{E}| = E_0 \) at the trap center. In the focal plane, \( \vec{B}_\mathrm{trap} \) is strongest on opposite sides of the tweezer and oriented perpendicular to the polarization direction (main text Fig.~\ref{fig:intro}c).

Such synthetic magnetic fields have been shown to induce decoherence~\protect\citesupp{thompson_coherence_2013}, as spatial gradients in \( \vec{B}_\mathrm{trap}(r) \) couple the qubit and motional degrees of freedom. The strength of this coupling depends on the orientation between \( \vec{B}_\mathrm{trap} \) and the applied bias field \( \vec{B}_\mathrm{0} \). When the tweezer polarization is aligned \emph{parallel} to \( \vec{B}_\mathrm{0} \), the synthetic field is orthogonal (\( \vec{B}_\mathrm{trap} \perp \vec{B}_\mathrm{0} \)), contributing only at second order. In this case, the decoherence is strongly suppressed as long as \( |\vec{B}_\mathrm{0}| \gg |\vec{B}_\mathrm{trap}| \). Conversely, when the tweezer polarization is \emph{perpendicular} to the bias field, \( \vec{B}_\mathrm{trap} \parallel \vec{B}_\mathrm{0} \), the two fields add directly and modify the effective qubit splitting, resulting in a small differential displacement between the trapping potentials of the two qubit states. This displacement can be expressed as
\begin{equation}
    \delta y = \frac{g_m\, \partial_y B_{\text{trap}}^{(x)}}{m \omega^2},
\end{equation}
where \( \omega \) is the trap frequency and \( m \) is the atomic mass. Notably, this shift $\delta y$ is only about 30 pm, which is much smaller than the ground-state wavepacket size \( \sigma_y = \sqrt{\hbar/(2m\omega)} = 31\,\mathrm{nm}\), with \( \delta y / \sigma_y \sim 10^{-3} \). Thus, even in the presence of an unsuppressed synthetic field, static trap-induced dephasing remains negligible.

\begin{figure*}
    \includegraphics[width=180mm]{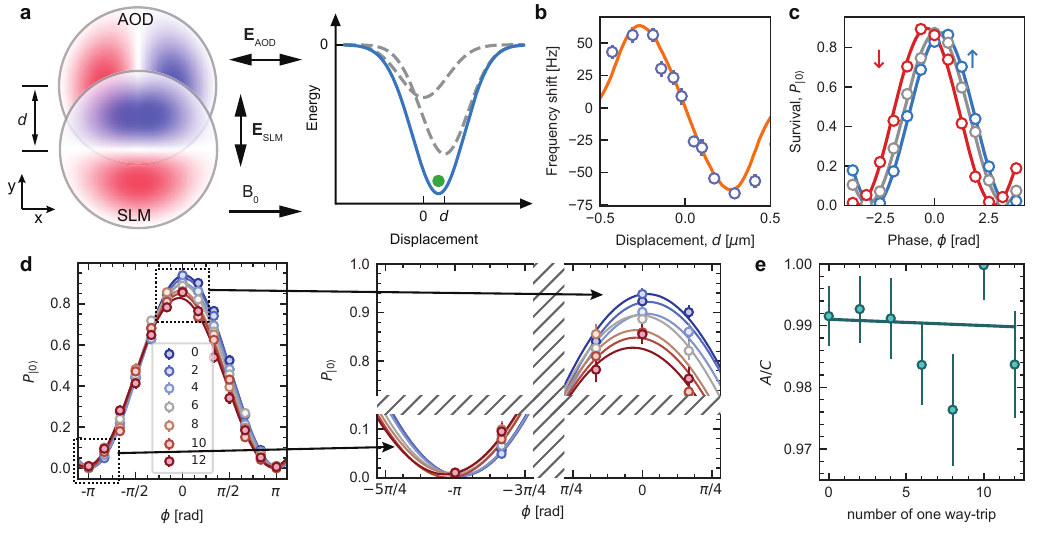}
    \caption{{\bf Decoherence during trap handoff.}
    (a) Schematic of the suboptimal configuration for SLM and AOD trap, with the two traps orthogonally polarized. The atom sees a summed trap potential of the two traps. When two traps are displaced in the $y$ direction, the atom is pulled away from the center of the SLM trap, experiencing an unsuppressed vector light shift.
    (b) Larmor frequency shift with various displacement $d$ shown in (a). The purple data is experimentally measured via Ramsey experiments, while the orange curve is the theoretical prediction with a single fitting parameter $\eta_{\mathrm{vls}}$. Based on this measurement, we estimate the $\eta_{\mathrm{vls}}=3\times 10^{-4}$ at a wavelength of 487 nm.
    (c) The Ramsey fringes after a round-trip between the storage and gate zones under $d=-240\, \mathrm{nm}$ (red), $d=0$ (gray) and $d=240\, \mathrm{nm}$ (blue).
    (d) Measurement of the Ramsey fringes with various number of one-way-trips that the atoms undergo during the Ramsey experiment (same dataset as the blue curve in Fig.~\ref{fig:intro}e).
    (e) Fitted contrast $A/C$ as a function of the number of one-way-trip. A linear fit gives a spin flip rate of $1(6)\times10^{-4}$ per one-way trip.}
    \label{fig:vls}
\end{figure*}

However, when traps are overlapped together, atoms may be pulled away from the center of one trap by another. In this case, if the synthetic field generated by any trap is parallel to $\vec{B}_\mathrm{0}$, they will directly shift the energy and the qubit accumulates a phase that depends on the relative alignment of the traps (Extended Data Fig.~\ref{fig:vls}a). This effect can be harnessed to probe the longitudinal polarization~\protect\citesupp{tomita_atom_2024} and extract \( \eta_\mathrm{vls} \) at our tweezer wavelength (Extended Data Fig.~\ref{fig:vls}b).

Moreover, this effect will contribute to decoherence during atom handoffs between traps in the case where SLM and AOD tweezers have orthogonal polarizations, as any misalignment in the direction transverse to \( \vec{B}_0 \) introduces an uncontrolled phase shift during transport. The phase shift is proportional to the displacement, and we measure it to be 7 mrad/nm per one-way trip between the storage and gate zones under this configuration (Extended Data Fig.~\ref{fig:vls}c). Together with the coherence measurement (main text Fig. \ref{fig:intro}e), we can estimate that we likely have a positional uncertainty of $80$ nm when aligning the SLM and AOD traps, combining both the site-to-site difference and fluctuation over time.

The optimal trap polarization configuration for preserving qubit coherence during transport is to align both the SLM and AOD trap polarizations parallel to the bias magnetic field $\vec{B}_0$. To quantify the coherence under this configuration, we perform a Ramsey experiment in which atoms are transported between the storage and gate zones. By measuring the Ramsey contrast as a function of the number of one-way trips, we observe that coherence remains unaffected by the transport, even in the absence of dynamical decoupling. Because atoms experience loss during transport, each Ramsey fringe is fit using a sinusoidal function of the form
\begin{equation}
    P_{\ket{0}} = A \cos(\phi - \phi_0) + C,
\end{equation}
where $\phi$ is the phase of the second $\pi/2$ pulse and $P_{\ket{0}}$ is the measured survival probability. The coherence is quantified by the normalized contrast $A/C$. Fitting this value as a function of the number of one-way trips yields an estimated decoherence rate of $1(6) \times 10^{-4}$ per trip (Extended Data Fig.~\ref{fig:vls}d and e).

\subsection*{Periodic trap modulation}
\label{sec:trapmodulation}

To mitigate inhomogeneous light shifts from optical tweezers during operations, the conventional approach is to turn off the trapping light during sensitive procedures such as entangling gate operations or state preparation~\protect\citesupp{bluvstein_quantum_2022, ma_high-fidelity_2023, evered_high-fidelity_2023, tsai_benchmarking_2025}. However, switching the trap on and off in an irregular manner leads to heating and atom loss, limiting the achievable circuit depth. As an alternative, we employ {\it periodic trap modulation}, in which a fixed frequency modulation of the trap is ramped on adiabatically and maintained for the duration of the experiment~\protect\citesupp{tiecke_nanophotonic_2014}. Provided the modulation is much faster than the trap frequency, this results in stable micromotion at the modulation frequency but does not heat the atom.

\begin{figure*}
    \includegraphics[width=180 mm]{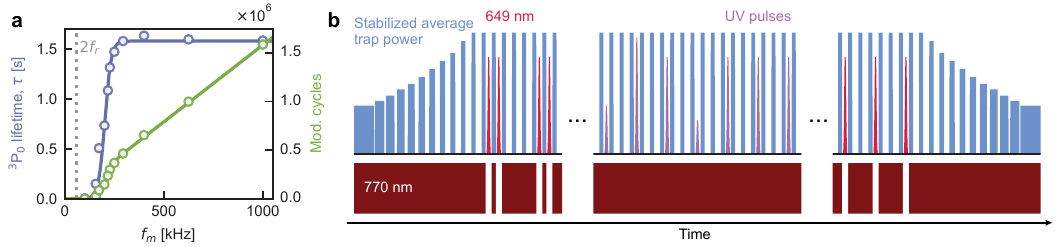}
    \caption{{\bf Periodic trap modulation.} (a) Measured lifetime $\tau$ of the \tpz qubit in an optical trap under different trap modulation frequencies $f_m$ (blue dots). Green dots show the corresponding number of modulation cycles $n = \tau f_m$, which exceeds $10^6$ for experimentally realistic $f_m$ values. (b) Schematics of synchronizing trap modulation with entangling gates and optical pumping. To switch on (off) trap modulation, we adiabatically ramp down (up) the duty cycle while keeping the averaged trap power unchanged.}
    \label{fig:trapmod}
\end{figure*}

We experimentally validate this approach using atoms in the $\mathrm{^3P_0}$ state. The trap is modulated with a 50\% duty cycle square wave at $f_m$ using an AOM, and the lifetime is measured. The average trap depth is stabilized at $V_\text{avg} = k_B\times\text{37}$ $\mu$K using photodiode feedback through a low-pass filter, corresponding to a trap frequency $f_r = \text{30}$ kHz in the radial direction. The lifetime begins increasing once the modulation frequency is higher than $2f_r$, but does not saturate at the scattering-limited lifetime until nearly $10f_r$ (Extended Data Fig.~\ref{fig:trapmod}a). When measuring the lifetime of the $^1S_0$ state in modulated traps (which is not limited by Raman scattering), we have observed lifetimes with large modulation frequencies that significantly exceed the lifetime without modulation, by a factor of 2. A similar effect was observed (but not reported) in Ref.~\protect\citesupp{tiecke_nanophotonic_2014}, and we do not currently have an explanation.

For operation of the circuits and benchmarking sequences in this work, we choose a trap modulation frequency of $f_m=400$ kHz. This trap-off window (1.3\,$\mu$s) is sufficient to accommodate our AR CZ gate pulse. The modulation is ramped on (or off) adiabatically over 100~$\mu$s by decreasing (or increasing) the duty cycle, while keeping the total trap power constant. At this modulation frequency, over 600,000 modulation cycles can be applied within the lifetime of the qubit, providing ample opportunity for applying operations. The timing scheme is illustrated in Extended Data Fig.~\ref{fig:trapmod}b. 

The periodic trap modulation scheme also improves qubit initialization and readout, both of which rely on the multi-photon optical pumping between the ground state and the metastable state. In particular, population will have to pass through the $\mathrm{^3P_2}$ state, which is anti-trapped at our tweezer wavelength. To prevent atom loss during this step, we synchronize the optical pumping pulses (649 nm pulses, driving the $\mathrm{^3P_0 \rightarrow {}^3S_1}$ transition) with the trap-off windows, as illustrated in Extended Data Fig.~\ref{fig:trapmod}b. This coordination ensures that atoms are never pumped into anti-trapped states while the trap is on, while also avoiding heating that would otherwise result from irregular trap switching. 

\subsection*{Design of amplitude-robust Controlled-Z gates}
\label{sec:arcz}

\begin{figure*}
    \includegraphics[width=180mm]{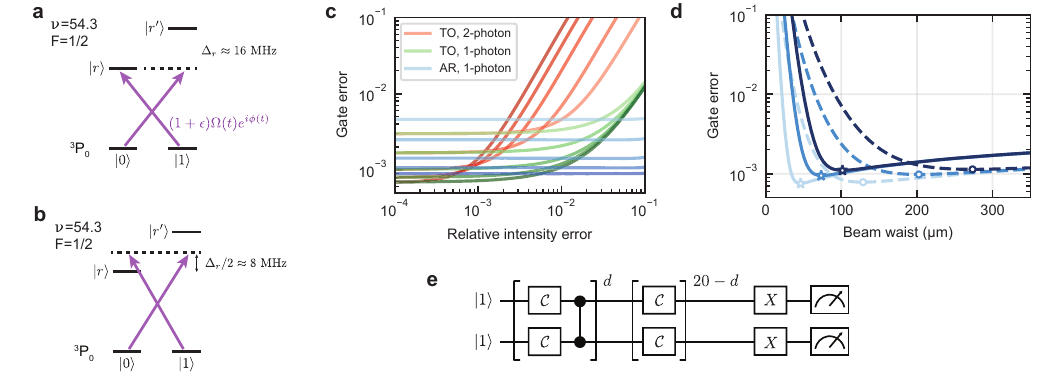}
    \caption{\textbf{AR CZ gate and local single-qubit phase gate.} (a) In our experiment, the Rydberg laser is linearly polarized perpendicular to the quantization axis, resulting in equal $\sigma^+$ and $\sigma^-$ components. The laser is tuned to the $\ket{1}$ to $\ket{r} = \ket{\nu = 54.3, F = 1/2, m_F = -1/2}$ resonance, but also couples $\ket{0}$ to $\ket{r'} = \ket{\nu = 54.3, F = 1/2, m_F = 1/2}$ with a detuning $\Delta_r$ from the Zeeman splitting of the Rydberg state.
    (b) For the local single-qubit phase gate, we tune the Rydberg beam to the midpoint between the two Zeeman-split Rydberg transitions, so that the $\sigma^+$ ($\sigma^-$) component off-resonantly couples $\ket{0}\!\rightarrow\!\ket{r'}$ ($\ket{1}\!\rightarrow\!\ket{r}$) with equal detuning magnitude $\Delta_r/2 \approx 8~\mathrm{MHz}$. The off-resonant coupling produces a differential AC Stark shift that implements $R_Z$ rotations.
    (c) Simulated gate infidelity (including spontaneous decay) versus relative laser-intensity error for the AR and TO gates, using Rydberg lifetime $\tau_{\rm Ryd}=88~\mu{\rm s}$ and the two-photon differential light-shift coefficient $\chi_{\rm 2-photon} \approx 0.945~$MHz$/$MHz$^2$ from Ref.~\protect\citesupp{evered_high-fidelity_2023}. Curves are shown for nominal Rabi frequencies  $\Omega/2\pi=$ 2, 4, 8, 12 and 16 MHz (light to dark). The AR gate is driven via a single-photon Rydberg transition and therefore has negligible intensity-dependent light shift. The TO gate is simulated for both single-photon and two-photon Rydberg implementations; the two-photon case includes the additional differential light shift on top of the spontaneous-decay contribution.
    (d) Simulated gate infidelity (including spontaneous decay), averaged over the targeted gate-zone, is plotted against the waist of an elliptical laser beam along the gate-zone direction. The curves correspond to targeted gate-zone sizes of 30~$\mu$m, 50~$\mu$m, and 70~$\mu$m, shown in colors ranging from light to dark, respectively. The total laser power is held constant at 3.75~W, and the waist perpendicular to the gate-zone direction is fixed at 12~$\mu$m (1/$e^2$ radius). Solid curves represent AR gates, and dashed curves represent TO gates without the light shift. The star and circle markers denote the respective optimal fidelities achieved for each gate type and target zone size.
    (e) Randomized circuit characterization of the CZ gate as a function of circuit depth $d$. The protocol follows that introduced in Ref.~\cite{peper_spectroscopy_2025}.}
    \label{fig:grape}
\end{figure*}

The AR CZ gate implemented in this experiment is designed using optimal control techniques following Ref. \protect\citesupp{jandura_optimizing_2023}. We extend that work by also considering off-resonant coupling from $\ket{0}$ to other $m_F$ sublevels of the Rydberg state (Extended Data Fig.~\ref{fig:grape}), and to incorporate the finite bandwidth of the AOM generating the laser pulse.

Under the assumption of a perfect Rydberg blockade, atom pairs initialized in $\ket{q} \in \{\ket{00}, \ket{01}, \ket{11}\}$ evolve disjoint subspaces. The time-dependent Hamiltonian for each initial state is given by:
\begin{align}
    H_{00}(t) &= \frac{(1 + \epsilon) \Omega (t) e^{i\phi(t)}}{2}(\ket{0r'}\bra{00} + \ket{r'0}\bra{00}) + \mathrm{h.c.} + \Delta_r (\ket{0r'}\bra{0r'} + \ket{r'0}\bra{r'0}), \\
    H_{01}(t) &= \frac{(1 + \epsilon) \Omega (t) e^{i\phi(t)}}{2}(\ket{0r}\bra{01} + \ket{r'1}\bra{01}) + \mathrm{h.c.} + \Delta_r \ket{r'1}\bra{r'1}, \\
    H_{11}(t) &= \frac{(1 + \epsilon) \Omega (t) e^{i\phi(t)}}{2}(\ket{1r}\bra{11} + \ket{r1}\bra{11}) + \mathrm{h.c.},
\end{align}
where we omit $\ket{10}$ due to its symmetry with $\ket{01}$. Here, $\Omega(t)$ and $\phi(t)$ denote the amplitude and phase profiles of the Rydberg laser, respectively, and $\epsilon$ represents quasi-static variations in laser intensity that we wish to be robust against.

For each initial state $\ket{q} \in \{\ket{00}, \ket{01}, \ket{11}\}$, the final state after gate evolution is given by
\begin{equation}
    \left(\mathcal{T}e^{\int_0^T i H_q(t) dt}\right)\ket{q} = \ket{\psi_q} = \ket{\psi_q^{(0)}} + \epsilon \ket{\psi_q^{(1)}} + \mathcal{O}(\epsilon^2),
\end{equation}
where $T$ is the gate duration. To realize a CZ gate, we require that the pulse satisfies $\ket{\psi_{q}^{(0)}} = e^{i\theta_q} \ket{q}$ and $\theta_{11} = 2 \theta_{01} - \theta_{00} + \pi$. By selecting an appropriate amplitude profile $\Omega(t)$ and Zeeman splitting $\Delta_r$, we can design the gate to be amplitude-robust by ensuring that $\ket{\psi_q}$ remains first-order insensitive to $\epsilon$, up to a global phase. This results in:
\begin{equation}
    \ket{\psi_q^{(1)}} = i\alpha\ket{\psi_q^{(0)}},
\end{equation}
where $\alpha = \left. \frac{\mathrm{d}\theta_{00}}{\mathrm{d}\epsilon} \right|_{\epsilon=0}$ is a real parameter that can be freely adjusted in GRAPE optimization. We note that since $\theta_{00}$ is non-zero, the amplitude-robust condition is different from the one used in Ref. \protect\citesupp{jandura_optimizing_2023}, where it simply requires $\ket{\psi_q^{(1)}}=0$.

The AR CZ gate is then obtained by minimizing the cost function:
\begin{equation}
    J = 1 - \mathcal{F} + \sum_q \left|\ket{\psi_q^{(1)}} - i\alpha\ket{\psi_q^{(0)}}\right|^2,
\end{equation}
where $\mathcal{F}$ represents the fidelity of the CZ gate at $\epsilon=0$, and $q$ is summed over $\{00, 01, 11\}$. To enable tractable optimization, the phase $\phi(t)$ is approximated as piecewise constant, $\phi_n$, over intervals $t \in [nT/N, (n+1)T/N)$ with $N=400$. To reduce rapid variations in $\phi(t)$ that may be challenging to implement experimentally, we additionally include a regularization term proportional to $\sum_{n=0}^{N-1}(\phi_{n+1} - \phi_n)^2$ in the cost function $J$. The amplitude profile $\Omega(t)$ is set to a constant $\Omega_0$, with sinusoidal rising and falling edges. The rise (fall) time is chosen to be much longer than $1/\Delta_r$ to suppress off-resonant oscillations between $\ket{0}$ and $\ket{r'}$ at the beginning (end) of the gate. Although this extends the total gate duration, it does not increase the time atoms spend in the Rydberg state. With the values of $\Delta_r/\Omega_0=6.4$ and $T=20.4\times\Omega_0^{-1}$, we obtain an AR CZ gate with simulated fidelity of $1-\mathcal{F} < 5 \times 10^{-4}$ in the absence of Rydberg decay.

As mentioned in the main text, the AR CZ gate reduces the frequency of recalibration in our experiment, and we use it in all logical-circuit demonstrations. Although its fidelity is lower than that of the TO gate at the nominal Rabi frequency, we emphasize here the broader applicability of the AR CZ gate—particularly in the high-Rabi-frequency regime. For blockade-based CZ gates, the dominant fundamental error arises from Rydberg-state decay; improving gate fidelity therefore requires increasing the Rabi frequency to shorten the gate duration. However, intensity fluctuations due to laser noise, pointing instability, or inhomogeneity across a gate zone impose a floor on the achievable error (Extended Data Fig. \ref{fig:grape}c). For gates implemented with two-photon excitation, this challenge is more severe because the AC Stark shift introduces an additional detuning error on top of the intensity fluctuations. This makes the requirements on laser-intensity stability and spatial homogeneity extremely stringent; for example, achieving a fidelity of 0.999 with the TO gate requires suppressing the relative intensity error below 0.08\% \protect\citesupp{evered_high-fidelity_2023}. These constraints are relaxed by eliminating the AC Stark shift using single-photon excitation, as demonstrated in this work. Furthermore, single-photon excitation enables the use of the robust AR gate, which further reduces sensitivity to intensity fluctuations and thus substantially relaxes stability and homogeneity requirements.

Moreover, for logical-circuit implementation, one generally prefers a wide gate zone to allow higher parallelism and support larger code distances~\protect\citesupp{bluvstein_logical_2024}. In such cases, balancing beam homogeneity against achievable Rabi frequency becomes essential. For an elliptical Gaussian beam with the beam waist along the gate-zone direction treated as a free tuning parameter, and under a fixed total laser power, the optimal zone-averaged fidelities of the AR and TO gates are nearly identical (0.99925 and 0.99922, respectively, for a 30-$\mu$m gate zone; Extended Data Fig.\ref{fig:grape}d), while the AR gate provides the additional robustness to laser noise and pointing. Therefore, we believe that the AR CZ gate is not only a practical choice for the present system, but also a scalable approach for reaching the high-fidelity regime in neutral-atom quantum processors.

We note that the lower TO gate fidelity reported in the main text compared to Ref.~\protect\citesupp{peper_spectroscopy_2025} is believed to result from excess laser phase noise, based on a decrease in the observed Rydberg $T_2^*$ time (from 5.7 $\mu$s to 3.5 $\mu$s).

\subsection*{Local single-qubit phase gate}
The single-qubit phase gate $R_Z(\theta)$ is implemented using an off-resonant Rydberg pulse applied while a single target atom is positioned in the Rydberg beam. As shown in Extended Data Fig.~\ref{fig:grape}b, the UV pulse is detuned by approximately $8~\mathrm{MHz}$, centered between the $\ket{1}\!\leftrightarrow\!\ket{r}$ and $\ket{0}\!\leftrightarrow\!\ket{r'}$ transitions, generating a differential light shift between $\ket{0}$ and $\ket{1}$. The pulse duration is fixed at $1~\mu\mathrm{s}$, and varying the pulse amplitude allows realization of arbitrary rotation angles $\theta$.

\addvspace{1ex}  
\subsection*{QEC circuit implementation}
\label{sec:code}

The $[[4,2,2]]$ code is a four-qubit error detection code with stabilizers $X_1X_2X_3X_4$ and $Z_1Z_2Z_3Z_4$. 
The logical states we demonstrate are Greenberger-Horne-Zeilinger (GHZ) states:
\begin{align}
    \ket{00}_L &= (\ket{0000}+\ket{1111})/\sqrt{2},\\
    \ket{++}_L = H^{\otimes 4}\ket{00}_L &=(\ket{\text{{+}{+}{+}{+}}}+\ket{{{-}{-}{-}{-}}})/\sqrt{2}.
\end{align}
More details of the code can be found in Ref.~\citesupp{grassl_codes_1997, linke_fault-tolerant_2017}. 

\begin{figure*}
    \centering
    \includegraphics[width= 180 mm]{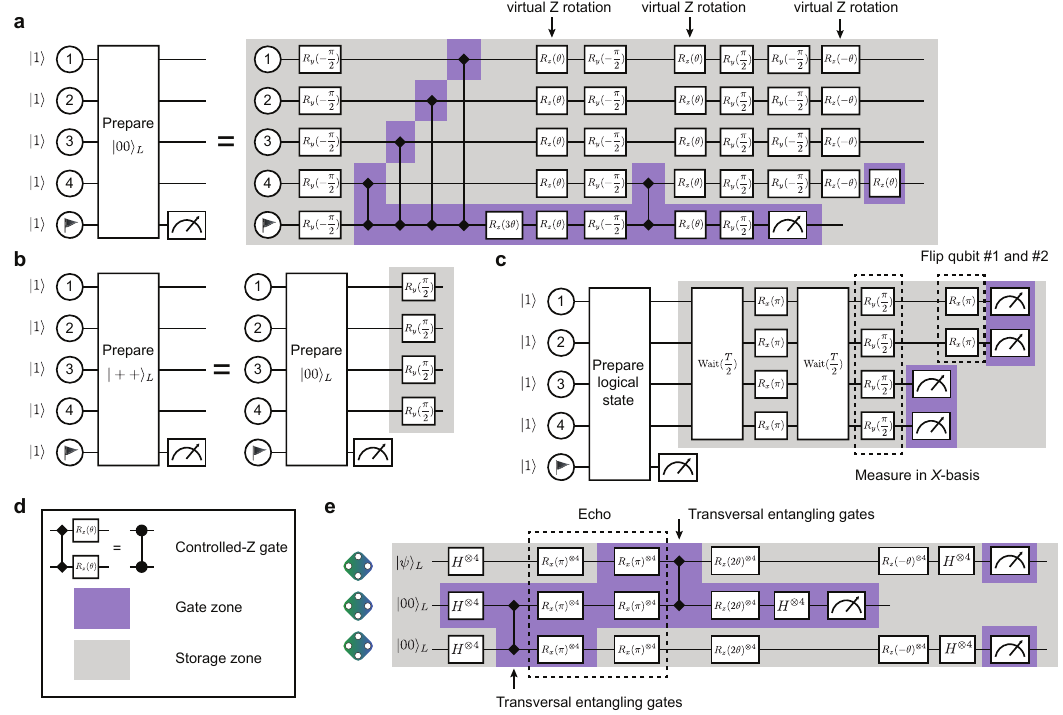}
    \caption{\textbf{Experimental implementation of quantum circuits.}
    (a) Encoding circuit for preparing logical qubit states $\ket{00}_L$ with the $[[4,2,2]]$ code. A flag qubit is used to herald certain Pauli errors and leakage errors during the preparation circuit.
    (b) $\ket{++}_L$ can be prepared by adding extra $R_y(\pi/2)$ gates on all data qubits after preparing $\ket{00}_L$.
    (c) Circuit for the ``encode-hold-decode" experiment. Logical qubits are measured directly in the $Z$-basis, or in the $X$-basis by inserting an additional $\pi/2$-pulse. During the wait, we implement a Hahn echo sequence to extend the coherence time. Before the final measurement, we
    apply an extra $\pi$ rotation on the first and second physical qubit to prevent the misidentification mentioned in the text.
    (d) The native entangling gates are a CZ gate up to a single qubit phase rotation $\theta\approx-3.4$ rad. Moreover, throughout this figure, we
    use purple (grey) shade to represent operations applied in the gate (storage) zones.
    (e) Teleportation circuit on distinct logical blocks encoded by the $[[4,2,2]]$ code. }
    \label{fig:full_circuits}
\end{figure*}

The encoding circuits for both states are presented in Extended Data Fig.~\ref{fig:full_circuits}a and b. 
The state preparation circuit is fault-tolerant to a single physical error (Pauli or atom loss/leakage) when postselecting on a flag qubit.
In the absence of errors, the flag qubit is expected to be in the bright state ($\ket{0}$) at the end of the preparation circuit.
The circuit is structured so that any high-weight errors originate only from the flag qubit.
Whether the single physical error is of Pauli or leakage type, it results in a dark readout on the flag qubit. 

Extended Data Fig.~\ref{fig:full_circuits}c shows the circuits for the ``encode-hold-decode" experiments. 
Since the GHZ states are four times more sensitive to dephasing than the physical qubits, a simple Hahn echo is applied to mitigate decoherence during the waiting period. 
The circuit depicted in the figure represents measurements in the $X$-basis. 
By omitting the final $R_y\left(\frac{\pi}{2}\right)^{\otimes 4}$ gates, measurements in the $Z$-basis can be performed. 
Outcomes with either all-dark or all-bright detections in the $Z$ ($X$) basis are decoded as the logical state $\ket{00}_L$ ($\ket{++}_L$).
One potential source of overestimating the decoding success rate arises when the atom loss rate is high, as four lost qubits, which appear dark, may be misinterpreted as the correct logical state. 
To prevent this misinterpretation, we intentionally flip the first and second qubit right before the final transverse measurement and instead expect an outcome of ``dark-dark-bright-bright" or ``bright-bright-dark-dark".

The $[[4,2,2]]$ code teleportation circuit is shown in Extended Data Fig.~\ref{fig:full_circuits}e. 
We interleave logical block transportation and two sets of $R_x(\pi)$ gates to cancel out a small magnetic field gradient between the gate and storage zones.

\subsection*{Stabilizer circuit simulation with erasure errors}
\label{sec:sim}

We develop a numerical model for simulating the quantum circuits implemented in this work, which is based on the standard stabilizer tableau method, with additional functionality to track qubits that experience leakage or erasure errors. Using independently characterized system parameters summarized in Extended Data Table~\ref{tab:benchmark}, we construct error models to realistically capture the behavior of our system. Below, we highlight a few notable aspects of the simulation treatments.

Any idling time $ t $ for metastable state qubits introduces a qubit leakage rate of $ p_{L} = 1 - \exp{(-t / \lt)} $. Under this leakage rate, the erasure fraction is measured to be approximately $ r_{e,l} \approx 0.72 $. The Ramsey fringe contrast indicates a physical $ T_2^* = 0.39 $ s, exhibiting a profile of Gaussian decay. Therefore, after the qubits spend $\tprep = 30$ ms on the equator of the Bloch sphere in the encoding circuit, we insert a $ Z $ error with a probability of $ 1 - \exp{[-(\tprep / T_2^*)^2]} $ on each physical qubit. During the waiting time $ \twait$, since a Hahn echo is applied, $T_2$ becomes the relevant timescale. Hence, we insert a $ Z $ error with a probability of $ 1 - \exp{(-\twait / T_2)}$, as it exhibits an exponential-decay profile.

The relaxation time $T_1 $ is much longer than $\tprep$ and more relevant to $\twait$ (up to 0.5 s) and also shows an exponential-decay profile, so we apply a bit-flip error with a probability of $ 1 - \exp{(-\twait / T_1)} $ on each physical qubit after the waiting time.

The error model for single-qubit RF gates is based on the single-qubit depolarizing channel, together with leakage out of the metastable manifolds and the same erasure fraction $r_{e,l}$. The total error rate of a single-qubit gate is measured to be $\esq = 1.0\times 10^{-3}$. Since a single-qubit gate takes $\tsq = 1.13$\ ms, the probability of leakage during a single-qubit gate is $ p_{L,\rm{1Q}} = 1 - \exp{(-\tsq / \lt)} = 7\times 10^{-4}$. The remaining error rate $\epsilon_{d,\rm{1Q}}=\esq-p_{L,\rm{1Q}}=3\times 10^{-4}$ is therefore contributed by the depolarizing channel.

The total error rate of the two-qubit entangling gate is measured to be $\ecz=0.016(1)$, the error model in simulation is a combination of Rydberg decay channel and depolarizing errors. Using the experimentally measured Rydberg decay branching ratios~\protect\citesupp{ma_high-fidelity_2023}, we reconstruct the decay process as consisting of three outcomes: erasure to the ground state ($^1$S$_0$, with erasure fraction $r_{e, \text{CZ}}$), Pauli-type errors to the metastable state ($^3$P$_0$), and leakage to other states. All remaining errors are modeled as a Pauli error. The Pauli error is expected to be predominantly $Z$-type. This is because the dominant sources of error---such as Doppler shifts, laser phase noise, and control imperfections---do not induce population transfer between $\ket{0}$ and $\ket{1}$. Additionally, while the fidelity of the local phase gate is not directly measured, we model it using a noise channel with an error rate comparable to that of the CZ gate, since it is also mediated by the Rydberg laser.

The qubit state preparation, measurement, and mid-circuit erasure detection methods are detailed in \textit{Experimental implementation}. Qubit initialization is performed through optical pumping (OP), and spin-state readout utilizes Rydberg transition and auto-ionization (RO), resulting in a combined state preparation and measurement error rate primarily due to atom loss, quantified as $\epsilon_{\text{OP}} + \epsilon_{\text{RO}} = 0.009$. Terminal qubit imaging is implemented via the $^1\text{S}_0$ to $^3\text{P}_1$ (556 nm) transition, with a false positive rate of $\epsilon_{\text{term,FP}} = 0.001$ and a false negative rate of $\epsilon_{\text{term,FN}} = 0.005$. Mid-circuit erasure detection is accomplished through the $^1\text{S}_0$ to $^1\text{P}_1$ (399 nm) transition, characterized by a false positive rate of $\epsilon_{\text{ed,FP}} = 0.014$ and a false negative rate of $\epsilon_{\text{ed,FN}} = 0.014$.

\vspace{10pt}
\textbf{Data Availability} The source data underlying the figures in this study have been deposited in Zenodo and are publicly available at https://doi.org/10.5281/zenodo.19491381.

\clearpage

\begin{table}
\centering
\begin{tabular}{l | r}
    \hline\hline
    Metric & Value \\
    \hline
    Optical pumping error (${}^1\text{S}_0\leftrightarrow{}^3\text{P}_0$ round trip) $\epsilon_{\text{OP}}$ & $0.60(3)\%$ \\
    Spin readout error $\epsilon_\text{RO}$ & $0.3(2)$\% \\
    Leakage error during transportation & $0.5(1)\%$ \\
    Dephasing error per storage-gate zone trip & $0.2$\% \\
    Single-qubit Clifford gate error $\esq$ & $0.10(1)\%$ \\
    Single-qubit Clifford erasure fraction $r_{e, \text{SQ}}$ & $\sim 0.56$ \\
    AR CZ gate error $\ecz$ & $1.7\%$\\
    AR CZ erasure fraction $r_{e, \text{CZ}}$ & $\sim 0.38$ \\
    $T_2^*$ (Gaussian) & 0.39(1)\ s\\
    $T_1$ (exponential)& 13\ s\\
    $T_2$ (exponential)& 6(2)\ s\\
    ${}^3\text{P}_0$ lifetime $\lt$ (exponential)& 1.64(3)\ s\\
    Erasure fraction while idling $r_{e, \text{idle}}$ & $\sim0.72$\\
    Erasure fraction while moving $r_{e, \text{move}}$ & $\sim0.5$\\
    \hline\hline
\end{tabular}
\caption{Summary of parameters used in the numerical simulation, extracted from independent benchmarking experiments. Note that the transport dephasing error is specified for the sub-optimal polarization configuration used in the QEC circuits.}
\label{tab:benchmark}
\end{table}

\makeatletter
\begingroup
\let\temp@auxout\@auxout
\let\addtocontents\mb@addtocontents
\let\@auxout\@auxoutsupp
\let\jobname\@auxoutsuppname
\let\refname\refnamesupp
\let\bibname\refnamesupp
\let\usecounter\@newusecounter
\makeatother

\makeatletter
\endgroup
\makeatother

\end{document}